\documentclass[10pt,aps,prl,twocolumn,superscriptaddress,numerical]{revtex4-2}
\usepackage[utf8]{inputenc}
\usepackage[english]{babel}
\usepackage[T1]{fontenc}
\usepackage{MnSymbol}
\usepackage{amsmath,dsfont,amsfonts,bbm,pifont}
\usepackage[colorlinks,
citecolor=purple,
linkcolor=purple,
urlcolor=purple]{hyperref}
\usepackage{tikz}
\usepackage{lipsum}
\usepackage{dsfont}
\usepackage{physics}
\usepackage{braket}
\usepackage{orcidlink}
\usepackage{multirow}
\usepackage[export]{adjustbox}
\usepackage[dvipsnames,svgnames]{xcolor}
\usepackage{color}
\usepackage{booktabs}

\graphicspath{{images}}

\newcommand{\Haar}{{\text{Haar}}} % Haar measure
\newcommand{\Wg}{{\text{Wg}}} % Weingarten functions
\newcommand{\Ic}{I_\mathrm{\rm c}} % Coherent info
\newcommand{\Hol}{\chi} % Holevo info
\newcommand{\Noise}{\mathcal{N}}
\newcommand{\fidel}{f_2} % Noise channel
\newcommand{\idp}{e} % Identity
\newcommand{\swp}{s} % Swap
\newcommand{\Id}{\mathbbm{1}}
\newcommand{\PauliX}{{\text{X}}}
\newcommand{\PauliY}{{\text{Y}}}
\newcommand{\PauliZ}{{\text{Z}}}

\newcommand{\Ex}{\mathbb{E}} % Expected value
\newcommand{\nreplicas}{2}

\usepackage{makecell}

\renewcommand{\section}[1]{\emph{#1}--\!}

\begin{document}

\title{Error-Correction Transitions in Finite-Depth Quantum Channels}

\author{Arman Sauliere~\orcidlink{0000-0002-1778-7263}}
\affiliation{Laboratoire de Physique Th\'eorique et Mod\'elisation, CNRS UMR 8089, CY Cergy Paris Universit\'e, 95302 Cergy-Pontoise Cedex, France}
\affiliation{JEIP, UAR 3573 CNRS, Collège de France, PSL Research University,
11 Place Marcelin Berthelot, 75321 Paris Cedex 05, France}

\author{Guglielmo Lami~\orcidlink{0000-0002-1778-7263}}
\affiliation{Laboratoire de Physique Th\'eorique et Mod\'elisation, CNRS UMR 8089, CY Cergy Paris Universit\'e, 95302 Cergy-Pontoise Cedex, France}
\affiliation{JEIP, UAR 3573 CNRS, Collège de France, PSL Research University,
11 Place Marcelin Berthelot, 75321 Paris Cedex 05, France}

\author{Pedro Ribeiro~\orcidlink{0000-0001-7630-2054}}
\affiliation{CeFEMA-LaPMET, Departamento de Física, Instituto Superior Técnico, Universidade de Lisboa, Avenida Rovisco Pais, 1049-001 Lisboa, Portugal}
\affiliation{Laboratoire de Physique Th\'eorique et Mod\'elisation, CNRS UMR 8089, CY Cergy Paris Universit\'e, 95302 Cergy-Pontoise Cedex, France}
\affiliation{CY Advanced Studies, Maison internationale de la recherche, CY Cergy Paris Universit\'e, 95302 Cergy-Pontoise Cedex, France}

\author{Andrea De Luca~\orcidlink{0000-0003-0272-5083}}
\affiliation{Laboratoire de Physique de l'\'Ecole Normale Sup\'erieure, ENS, Universit\'e PSL, CNRS, Sorbonne Universit\'e, Universit\'e Paris-Cité, 75005 Paris, France}

\author{Jacopo De Nardis~\orcidlink{0000-0001-7877-0329}}
\affiliation{Laboratoire de Physique Th\'eorique et Mod\'elisation, CNRS UMR 8089, CY Cergy Paris Universit\'e, 95302 Cergy-Pontoise Cedex, France}
\affiliation{JEIP, UAR 3573 CNRS, Collège de France, PSL Research University,
11 Place Marcelin Berthelot, 75321 Paris Cedex 05, France}

\begin{abstract}
We study error-correction protocols in which a quantum channel encodes logical information into an enlarged Hilbert space through a one-dimensional random circuit with local gates. We consider both noise acting after the encoding step and noise affecting the encoding circuit itself. Using the coherent information, we show that in both settings the infinite-depth limit is governed by a universal random-matrix transition at a critical noise rate, separating a recoverable phase from one where information is lost. We then characterize the leading finite-depth deviations from this universal regime. When noise acts only after a unitary encoder, convergence to perfect encoding is exponential in circuit depth, although boundary effects can delay saturation beyond the design time. When the encoder itself is noisy, the relevant control parameter becomes the circuit fidelity, which replaces the Hashing bound, and convergence is only polynomial in depth.
\end{abstract}

\maketitle
\section{Introduction} Quantum computers are expected to unlock capabilities beyond those of classical devices, from quantum many-body simulation to algorithmic speedups in specific tasks~\cite{Feynman1982Simulating,Lloyd1996UniversalSimulators,Shor1997Algorithms,Grover1997Search,BernsteinVazirani1997QCT,AaronsonArkhipov2011LinearOptics,Zhong2020PhotonicAdvantage,Kretschmer2025}. Realizing this promise requires going beyond the noisy intermediate-scale quantum regime and entering the fault-tolerant one, where logical information can be processed reliably despite decoherence and imperfect control~\cite{Preskill2018NISQ,Quantinuum2025,Aasen2025,Peham2025}. The standard route is quantum error correction, which encodes logical degrees of freedom into a larger Hilbert space using structured codes~\cite{shor1995,gottesman1997stabilizercodesquantumerror,gottesman1998theory,Nielsen_Chuang_2010,Bluvstein2023,Gidney2021stimfaststabilizer}. Understanding the limits of such protection remains a central problem in quantum information science~\cite{YuanCowtanHeLinWilliamson2026ParsimoniousQLDPCSurgery,GuLiuQuintavalleXuEisertRoffe2026QGPU,WillsLinZhangHsieh2026LinearTimeQECC,PerlinHeArmenakasAndresMartinezHaoHermanJinMayerSelfAmaroRyanAndersonShaydulin2026FTExecution,HetenyiBrownWilliamson2026ConstantDepthCultivation,LoLyonsVerresenVishwanathTantivasadakarn2025S3Pedagogical,JingSalaJiangVerresen2025IntrinsicHeralding,SalaVerresen2025StabilityLoopModels,KnillLaflammeViola2000GeneralNoiseQEC,Wilde2013QuantumInformationTheory,Preskill2018Physics219}.

\begin{figure}
    \centering
    \includegraphics[width=\linewidth]{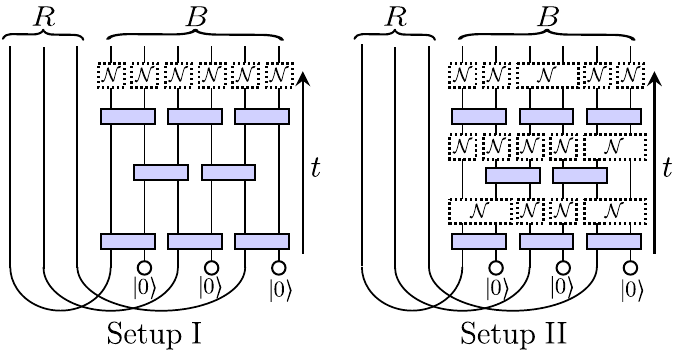}
    \makeatletter\long\def\@ifdim#1#2#3{#2}\makeatother
    \caption{We study the coherent information of a random encoding circuit at finite depth $t$ and investigate its ability to protect quantum information against noise, whether the noise acts after the encoder or within the encoding circuit itself. While Setup I is restricted to independent single-qudit noise, Setup II additionally includes multiqudit noise. We denote by $\gamma$ the error rate associated with each noise channel.}
    \label{fig:sketch_noisy_circuit}
\end{figure}

Alongside this structured paradigm, it is useful to study \emph{generic} encoding channels generated by random circuits~\cite{HaydenPreskill2007Mirrors,Choi2020_QEC_Scrambling_MeasurementInducedTransition,FanVijayVishwanathYou2020SOEC,TurkeshiSierant2024PRL,Sommers2025,Nelson2025_ErrorCorrectionPhaseTransition_NoisyRandomCircuits,DiasPerkovicHaqueRibeiroMcClarty2022NoiseSymmetryBreaking,YiLiuLi2025LovaszAQEC,KrollHelsen2026BrickworkClifford,LiShuZhu2025AQIM,trigueros2026nonstabilizernesserrorresiliencenoisy}. Here we study channels that encode $k=rN$ logical qudits into $N$ physical qudits through a one-dimensional brick-wall circuit of depth $t$ (Fig.~1), and we use the \emph{coherent information} $\Ic$ as our diagnostic of information transmission~\cite{DevetakShor2005,HaydenHorodeckiWinterYard2008}. Our goal is to determine how finite-depth random encoders develop error-correction thresholds as functions of depth and noise strength. We consider two scenarios (Fig.~\ref{fig:sketch_noisy_circuit}). In setup I, the encoding circuit is unitary and noise acts only after encoding on the physical qudits. In setup II, the encoding itself is noisy, with a noise channel applied after each unitary gate.

In setup I, the large-depth limit is governed by \emph{random-matrix} (RM) universality: the encoder generates an Haar-random code subspace, and the capacity problem reduces to an RM calculation closely related to decoupling~\cite{HaydenHorodeckiWinterYard2008}. One then finds a sharp transition at a critical noise rate fixed by the Hashing bound~\cite{Wilde2013QuantumInformationTheory}, $H=1-r$, separating an error-correcting phase from a lossy one. In the protected phase, recovery becomes essentially perfect, with $\Ic=k$ up to corrections exponentially small in $N$. These connections between capacity, scrambling, and open-system information dynamics have been widely explored~\cite{GullansHuse2020Purification,Choi2020_QEC_Scrambling_MeasurementInducedTransition,Zhuang2023_DynamicalPhaseTransitions_InformationFlow_RandomCircuits,Sommers2025}.

A main goal of this work is to go beyond this asymptotic RM picture and characterize the finite-depth corrections. In setup I, within the protected phase, the approach to perfect encoding is exponentially fast in depth, but with corrections of the form $N e^{-2t/\tau}+O(e^{-t/\tau})$, analogous to those associated with the formation of a quantum state design, governed by $N e^{-2t/\tau}$~\cite{lami2025anticoncentration,dowling2025freeindependenceunitarydesign,HarrowLow2009Designs,BrandaoHarrowHorodecki2016CMP,7rzk-2jyh}, where $\tau$ is the purity-decay timescale~\cite{Nahum2017EntanglementGrowthPRX,Nahum2018OperatorSpreadingPRX}.

In setup II, recently considered also in Ref.~\cite{Nelson2025_ErrorCorrectionPhaseTransition_NoisyRandomCircuits}, the encoder does not generically reach the same ideal RM universality at large depth. Nevertheless, we show that the coherent information retains the same universal functional form when expressed in terms of the \emph{circuit fidelity} $F$, whose logarithm replaces the Hashing bound as the relevant measure of noise. For circuits with fixed fidelity, the convergence to perfect encoding is much slower than in the previous case, with corrections inversely proportional to the circuit depth.

\section{Background} A quantum error-correcting code maps $k$ logical qudits into $N$ physical qudits through an encoding channel. The $k$ logical qudits, each of local dimension $d$, are supplemented by $N-k$ ancillas initialized in $|0\rangle$, and a quantum channel $\Lambda$ describing the encoding circuit and possible noise act on the full system. We denote the resulting physical output by $B$. Equivalently, the logical input can be taken maximally entangled with a reference system $R$ of $k$ qudits via Bell pairs $\ket{\phi_+}=d^{-1/2}\sum_{i=0}^{d-1}\ket{i}\ket{i}$, while $R$ remains untouched by $\Lambda$. The final state is then
\[
\rho_{RB}=(\Id_R\otimes \Lambda)(\ket{\pmb{\phi_+}}\bra{\pmb{\phi_+}}\otimes \ket{\pmb{0}}\bra{\pmb{0}}),
\]
where $\ket{\pmb{\phi_+}}=\ket{\phi_+}^{\otimes k}$ and $\ket{\pmb{0}}=\ket{0}^{\otimes N-k}$. The coherent information is
\begin{equation}\label{eq:Ic_def}
    \Ic(\Lambda)=S(\rho_{B})-S(\rho_{RB})\, ,
\end{equation}
where $S(\rho)$ denotes the Von Neumann entropy. The coherent information optimized over the input state for single-qudit noise $\mathcal{N}$ is related to the quantum channel capacity through
$\mathcal{Q}(\Noise)=\lim_{N \to \infty } 1/N \, \max_\rho \Ic(\rho,\Noise^{\otimes N})$~\cite{PhysRevA.55.1613,DevetakShor2005}. In our setting, the input state is chosen to be maximally mixed, i.e.,\ $\rho=\Id/d^k$. A value $I_{\mathrm c}(\Lambda)/k=1$ corresponds to perfect protection of the logical information, so a decoding circuit can fully recover the input state.

In this Letter, channel $\Lambda$ takes the form of a random quantum circuit of depth $t$, with local noise channels applied either after the circuit (setup I) or within the circuit itself (setup II). These two scenarios are illustrated in Fig.~\ref{fig:sketch_noisy_circuit}. In both setups, we consider the coherent information averaged over random circuit realizations. We denote by $U$ the noiseless unitary circuit, by $\Lambda_U$ the corresponding noisy quantum channel, and by $\mathbb{E}_U[\cdots]$ the average over circuit realizations. For analytical tractability, we make two standard simplifications: we replace the von Neumann entropy by the second Rényi entropy, and we compute the annealed average
\begin{equation}\label{eq:annealed_average}
        \Ic=-\log_d\Ex_{U}\Tr(\rho_B^\nreplicas)+\log_d\Ex_{U}\Tr(\rho_{RB}^\nreplicas).
\end{equation}
We focus on circuits whose unitary gates are independently drawn from the Haar measure. The averages over two replicas can then be evaluated using Weingarten calculus~\cite{Collins2003MomentsUnitaryGroup,CollinsSniady2006IntegrationClassicalGroups,CollinsMatsumotoNovak2022WeingartenCalculus}, which rewrites them in terms of permutation operators $\sigma\in S_{\nreplicas}$~\cite{Nahum2017EntanglementGrowthPRX,Nahum2018OperatorSpreadingPRX,chan2021manybody,Kostenberger_2021,Turkeshi2025,sierant2023entanglement}. In vectorized notation $O\to|O\rangle\!\rangle$, the basic formula is
\begin{equation}\label{eq:weingarten_formula}
\mathbb{E}_{U \sim \Haar(q)} \left[ (U \otimes U^*)^{\otimes \nreplicas} \right]
= \sum_{\pi, \sigma \in S_\nreplicas} \Wg_{\pi, \sigma}(q)\, |\pi\rangle\!\rangle \langle\!\langle \sigma| \, .
\end{equation}
The Weingarten matrix $\Wg(q)$ is the pseudoinverse of the overlap matrix $\left[G(q)\right]_{\pi,\sigma}:=\llangle \pi |\sigma\rrangle$. For a $q$-dimensional Hilbert space, the identity $\idp$ and swap $s$ permutations read as
\begin{equation}\label{eq:sigma_expanded}
   |\idp\rangle\!\rangle = \sum_{i, j = 0}^{q-1} |i i \rangle|j j\rangle, 
   \qquad
   |s\rangle\!\rangle = \sum_{i, j = 0}^{q-1} |i j \rangle|j i\rangle \, .
\end{equation}
Noise is modeled as a generic quantum channel $\Noise(\rho)=\sum_\alpha K_\alpha \rho K_\alpha^\dag$,
with normalized Kraus operators $\sum_\alpha K_\alpha^\dag K_\alpha=\Id$~\cite{Kraus1983StatesEffectsOperations,Wilde2013QuantumInformationTheory,Watrous2018TheoryQuantumInformation}. In the replicated observables of interest, the noise channel is always contracted with permutation states,  defining a modified noisy overlap matrix
\begin{equation}
\label{eq:Noisy gram}
    [\Tilde{G}(q,\Noise)]_{\pi,\sigma}:=\llangle \pi |\Noise^{\otimes \nreplicas}|\sigma\rrangle \,.
\end{equation}
For the single-qudit depolarizing channel $\Noise(\rho)=(1-\gamma)\rho+\gamma \Id/d$, this matrix becomes $\Tilde{G}(d,\Noise)=\bigl(\begin{smallmatrix} d^2 & d \\ d & 1+(d^2-1)(1-\gamma)^2 \end{smallmatrix}\bigr)$. We now analyze separately setup I (perfect encoding followed by noise) and setup II (noisy encoding).\\

\section{Setup I} In this case, the unitary encoding circuit $U$ is followed by single-qudit noise channels (see Fig.~\ref{fig:sketch_noisy_circuit}, left), so $\Lambda_{U}|\rho\rrangle=\Noise^{\otimes N} (U\otimes U^*)|\rho\rrangle$. The two averages appearing in Eq.~\eqref{eq:annealed_average} differ only in the boundary condition $|v_R\rrangle$ imposed on the system $R$. Specifically, the boundary term is $|v_B\rrangle=d^{-\nreplicas k}| \idp\rrangle$ for the entropy of $B$, and $|v_{RB}\rrangle=d^{-\nreplicas k}| s\rrangle$ for the entropy of $RB$. By contrast, both terms share the same boundary condition on system $B$, namely $\llangle v_0|=\llangle s|\mathcal{N}^{\otimes \nreplicas}$. Together, these boundary vectors give a unified expression for the two averages in Eq.~\eqref{eq:annealed_average}:
\begin{equation}
\label{eq:statistical_model}
\begin{split}
    &\Ex_{U}\Tr(\rho_{B / RB}^\nreplicas)=\llangle v_0|\Ex_U[(U\otimes U^*)^{\otimes \nreplicas}] |v_{B / RB}\rrangle\otimes |\pmb{0},\pmb{0}\rrangle^{\otimes 2}\,.
\end{split}
\end{equation}

We first consider the simple case in which $U$ is drawn from the Haar measure on the full Hilbert space, i.e.\ a global random unitary, which yields the RM prediction.
Using Eq.~\eqref{eq:weingarten_formula}, one obtains two independent sums over the symmetric group $S_\nreplicas$. Since all subsystem sizes scale extensively with 
$N$, in the large-$N$ limit the dominant contributions arise only from the configurations with $\sigma=\pi=\idp$ and $\sigma=\pi=\swp$. This leads to
\begin{equation}\label{eq:statistical_model_2}
\begin{split}
    \Ex_{U}\Tr(\rho_{B / RB}^\nreplicas)=d^{-\nreplicas N}\big[&\Tilde{G}_{\swp,\idp}(d,\Noise)^N \llangle e|v_{B/RB}\rrangle\\ &+\Tilde{G}_{\swp,\swp}(d,\Noise)^N \llangle s |v_{B/RB}\rrangle\big]\,,
\end{split}
\end{equation}
where we used the definition in Eq.~\eqref{eq:Noisy gram}. To simplify the notation, we define 
\begin{equation}
    g_{\pi,\sigma}=\log_d G_{\pi,\sigma}(d)-\log_d \Tilde{G}_{\pi,\sigma}(d,\Noise)\, .
\label{eq:def_g_12}
\end{equation}

Simplifying in the two cases, we find
\begin{equation}
\label{eq:RM purities}
    \begin{split}
        \Ex_{U}\Tr(\rho_B^\nreplicas)&=d^{-(g_{\swp,\idp}+1) N}+ d^{-(g_{\swp,\swp}+r)N}\\
        \Ex_{U}\Tr(\rho_{RB}^\nreplicas)&=d^{-(g_{\swp,\idp}+1+r)N}+d^{-g_{\swp,\swp} N}\,,
        %\Ex_{U}\Tr(\Lambda(\ket{\pmb{0}}\bra{\pmb{0}})^\nreplicas)&=d^{-(g_{\swp,\idp}+1) N}+d^{-g_{\swp,\swp} N}\,,
    \end{split}
\end{equation}
where we used $k=rN$. In the large-$N$ limit, this gives
\begin{equation}
\begin{split}
    \Ic/N &= \min( g_{\swp,\idp}+1,g_{\swp,\swp}+r)-\min(g_{\swp,\idp}+1+r,g_{\swp,\swp}) \,.
    %\Hol/N&= \min( g_{\swp,\idp}+1,g_{\swp,\swp}+r)-\min(g_{\swp,\idp}+1,g_{\swp,\swp})\, .
\end{split}
\label{eq:RM info}
\end{equation}

In the absence of noise, one has $g_{\swp,\idp}=g_{\swp,\swp}=0$, which yields $\Ic=k$. Moreover, the condition $g_{\swp,\idp}=0$ holds for any unital noise channel. In this regime, the information encoded in the $k$ logical qudits can be recovered perfectly. This behavior breaks down beyond a critical noise strength, determined by the condition $g_{\swp,\swp}-g_{\swp,\idp}=1-r$. When satisfied, the inequality $H_2\leq 1-r$, with $H_2:=g_{\swp,\swp}-g_{\swp,\idp}$, ensures the possibility of perfect recovery. This inequality is known as the \emph{Hashing bound}~\cite{Wilde2013QuantumInformationTheory}. For single-qubit Pauli channels $\Noise(\rho) = p_0 \Id + p_1 \PauliX \rho \PauliX + p_2 \PauliY \rho \PauliY + p_3 \PauliZ \rho \PauliZ$, the quantity $H_2$ reduces to the $2$-Rényi entropy of the probability vector $(p_0, p_1, p_2, p_3)$ (see End Matter for details)~\cite{Wilde2013QuantumInformationTheory}. \\

A useful perspective on Eq.~\eqref{eq:statistical_model_2} is to interpret it as the \emph{partition function} of an effective statistical mechanical model for a single spin, whose degree of freedom corresponds to the two possible states, identity $e$ and swap $s$, analogous to the $+$ and $-$ states of a standard Ising variable. This approach can be generalized beyond the simple case of a single global Haar matrix to brickwork random circuits, resulting in the partition function of a two-dimensional (2D) Ising-like model. Taking this perspective, and in the absence of noise, the term $\Ex_{U} \Tr(\rho_B^2)$ can be schematically represented as
\begin{equation}\label{eq:schemino1}
    \Ex_{U} \Tr(\rho_B^2)=\includegraphics[width=0.3\linewidth,valign=c]{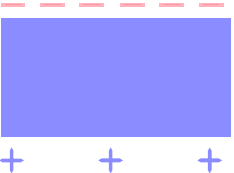}
    +\includegraphics[width=0.3\linewidth,valign=c]{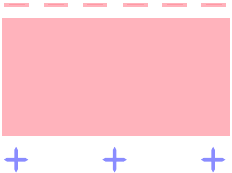}\,,
\end{equation}
where blue represents the identity $e$ permutation and red is the swap $s$. We use $+$ and $-$ signs at the boundaries to specify the boundary conditions, which, analogous to external magnetic fields in Ising models, favor either the identity ($+$) or the swap ($-$) state. When no sign is present, the boundaries are free, meaning that neither state is preferred. As shown in Eq.~\eqref{eq:schemino1}, in the absence of noise, the top boundary condition favors the $s$ permutation on all sites, whereas the bottom boundary condition favors the $e$ permutation. This latter effect appears when computing the purity of $B$ because the reference qudits in $R$ are traced out. Evaluating Eq.~\eqref{eq:schemino1} reduces to counting the boundary $+$ and $-$ signs that do not match the bulk state. This number is $N$ for the left configuration and $k$ for the right. Each mismatch contributes an exponential suppression factor, giving $\Ex_{U} \Tr(\rho_B^2) = d^{-N} + d^{-k}$, in agreement with the first line of Eq.~\eqref{eq:RM purities} for zero noise. Since $N \geq k$, the configuration with a swap permutation in the bulk is exponentially favored for large $N$. \\
However, upon introducing noise, the top boundary condition increasingly disfavors the $s$ permutation as the noise strength grows. We represent this effect by additional plus symbols at the output of the circuit, which encode the effective boundary field induced by the noise and increasingly favor the identity permutation.
\begin{equation}
\includegraphics[width=0.5\linewidth,valign=c]{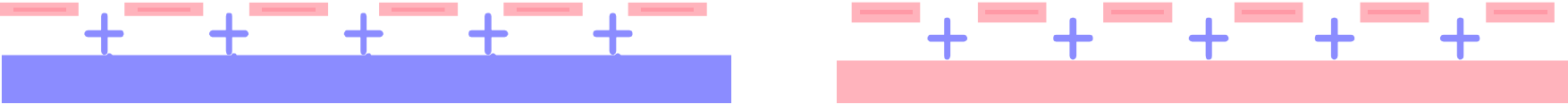}
\end{equation}
Each such symbol contributes a factor $d^{-g_{\swp,\idp}}$ for the left configuration and $d^{-g_{\swp,\swp}}$ for the right one. Since there are $N$ such contributions, we recover the result of Eq.~\eqref{eq:RM purities}. The critical point signaling the loss of perfect recovery corresponds to the point at which the identity and swap configurations become equally dominant.\\

\begin{figure*}[t!]
    \centering
    \includegraphics[width=\linewidth]{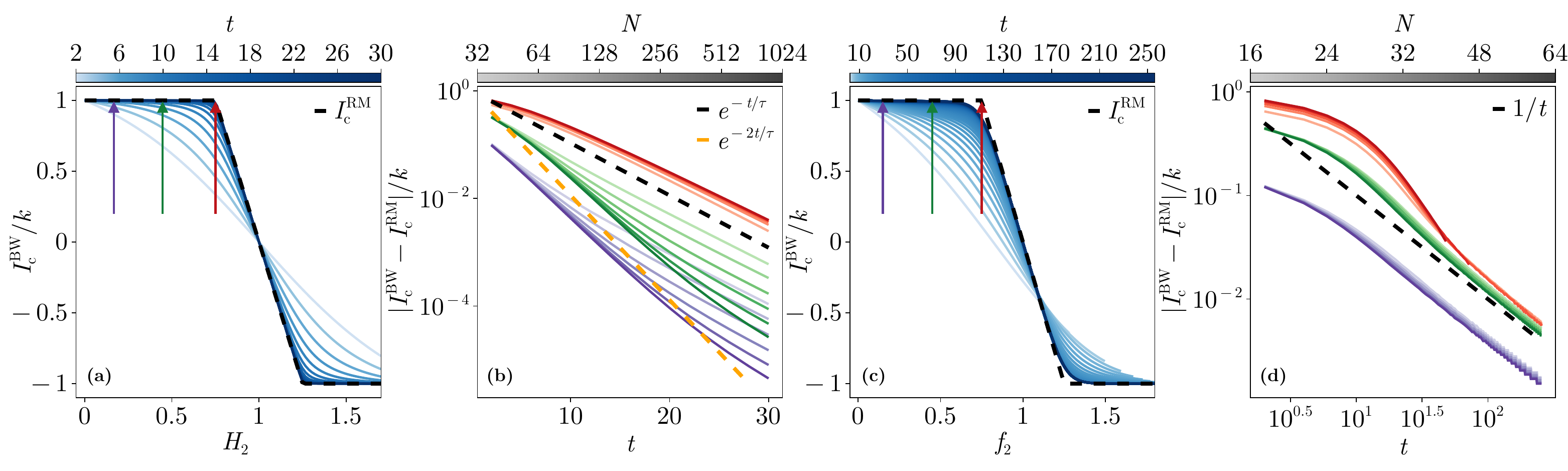}
    \caption{ Coherent information in setups I (\textbf{a},\textbf{b}) and II (\textbf{c},\textbf{d}). In both cases, we set the encoding rate to $r=1/4$ and use depolarizing noise channels (see Fig.~\ref{fig:info_ampdamp} for similar results with amplitude damping noise). Panel (a) shows the coherent information in setup I for system size $N=512$ as a function of the Hashing bound $H_2$. Panel (c) displays the coherent information for setup II with system size $N=32$ as a function of the parameter $\fidel$ [see Eq.~\eqref{eq:smallf}] and for different depths. In both panels (a) and (c) a range of values of the depolarizing strength $\gamma$ have been explored. Panels (b) and (d) display finite-time corrections to the RM result for setups I and II, respectively. Three distinct noise strengths, two within the recovery regime and one at the transition point, are considered [see arrows in panels (a) and (c)]. In setup I, panel (b), in the recovery regime, corrections initially decay as $N e^{-2t/\tau}$, the same scaling that governs the approach to a quantum state design, before crossing over to a slower decay $e^{-t/\tau}$; the early-time scaling is evidenced by the collapse of the corresponding curves. At the transition point, the corrections exhibit a scaling precisely of the form $N e^{-t/\tau}$, as demonstrated by the overlap of all curves. In setup II, panel (d), the corrections always follow a scaling proportional to $N/t$. For higher-replica numerical data see Fig.~\ref{fig:info_3_replicas}.}
    \label{fig:info_depolarizing}
\end{figure*}

% To characterize the finite-time corrections to the RM result, it is necessary to identify the classes of configurations that contribute at finite depth.  F
%or chaotic random Haar circuits $U$, Eq.~\eqref{eq:statistical_model}, obtained by repeatedly applying Eq.~\eqref{eq:weingarten_formula} to each gate of the circuit, 
We now consider the case where $U$ in Eq.~\eqref{eq:statistical_model} is a random quantum brickwork circuit of finite depth. Since each gate is independently drawn from the Haar measure, one can repeatedly apply Eq.~\eqref{eq:weingarten_formula}. This yields the partition function of a 2D Ising-like classical statistical model, with permutation degrees of freedom~\cite{FisherKhemaniNahumVijay2023RandomQuantumCircuits,Nahum2018}. The model features ferromagnetic interactions~\cite{FisherKhemaniNahumVijay2023RandomQuantumCircuits,Nahum2018}. The boundary vectors are the same of Eq.~\eqref{eq:statistical_model} and impose effective magnetic fields on the top and bottom edges. For shallow circuits $t =O( \log N)$, a simplified treatment of this model is possible. In fact, Renormalization Group arguments (see, for example, Ref.~\cite{sauliere2025universalityanticoncentrationchaoticquantum}) indicate that only configurations corresponding to \emph{vertical domains} survive, effectively reducing the problem to a simpler one-dimensional statistical-mechanics model. We illustrate the RG coarse-graining process as follows:
\begin{equation}\includegraphics[width=0.8\linewidth,valign=c]{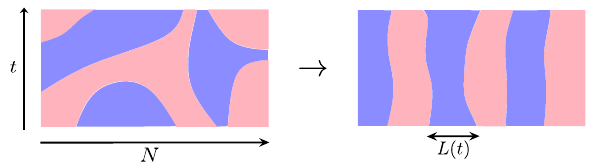}
\end{equation}
This procedure yields a characteristic length scale that controls the spatial extent of the vertical domains. This length scale is commonly referred to as the Thouless length $L(t)$  \cite{PhysRevLett.123.210603,PhysRevLett.130.140403,sauliere2025universalityanticoncentrationchaoticquantum}, $L(t)=L_0 e^{t/\tau}$, where the timescale $\tau$ is determined by the microscopic properties of the noiseless encoding circuit and controls the rate at which the half-chain purity decays, $P \propto e^{- t/\tau}$. For Haar-random brick-wall circuits, the scaling of the half-chain purity is explicitly known \cite{nahum2017quantum}, and is given by $\tau^{-1}=\log\left(\frac{d^2+1}{2d}\right)$. In the protected phase, in the large-time limit, the only contributing configuration is the uniform one with $s$ everywhere in the bulk. Spatially localized domains of $e$ can appear on top of this configuration. These domains may either be embedded within the bulk, be bounded by two domain walls, or occur at a boundary, where they are bounded by a single domain wall, namely
\begin{equation}\label{eq:domainwalls}
    \Ex_{U} \Tr(\rho_B^2)=\includegraphics[width=0.33\linewidth,valign=c]{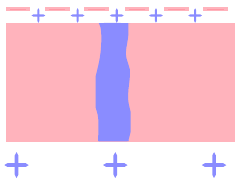}
    +\includegraphics[width=0.33\linewidth,valign=c]{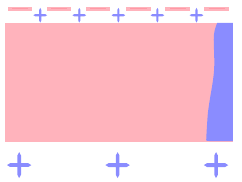}\,.
\end{equation}
Domains located in the bulk contribute a factor scaling as $N / [L(t)]^2$, where $N$ reflects the multiplicity, since they can appear anywhere along the $N$ qudits, and $1/L(t)$ is the cost of a vertical domain wall. By contrast, domains attached to a boundary are spatially constrained and consequently give a factor $1/L(t)$. At fixed $N$, this produces a crossover in the leading corrections to the RM result as time increases: initially they decay as $N e^{-2t/\tau}$ (coinciding with the decay of the frame potential to its Haar value, setting up the design time, see Fig.~\ref{fig:frame_potential}); however, once $L_0 e^{t/\tau}\gg N$, they cross over to a slower decay $e^{-t/\tau}$ (see Fig.~\ref{fig:info_depolarizing}, where we present numerical data for the coherent information for an encoding circuit followed by a single-qubit depolarizing channel). Finally, at the critical point $H_2(\gamma) = 1-r$, the identity and swap configurations acquire equal weight. As a result, a single domain wall can proliferate in the bulk, separating a blue and red region in Eq.~\eqref{eq:domainwalls}, with weight scaling as  
$N/L(t)$, which dominates over the two previously discussed contributions. \\

\section{Setup II} In this setting, due to the presence of noise in the circuit, it is important to first evaluate the average state fidelity of the output $\Lambda_{U}(\ket{\pmb{0}}\bra{\pmb{0}})$, where $\ket{\pmb{0}} = \ket{0}^{\otimes N}$, $\Lambda_{U}$ denotes a noisy circuit of $t$ layers, and $U$ is the noiseless unitary circuit. It can be written as $F=\Ex_{U} \left\{ \Tr\left[ \Lambda_{U}(\ket{\pmb{0}}\bra{\pmb{0}}) \, U(\ket{\pmb{0}}\bra{\pmb{0}})U^\dag \right] \right\}$. Using the swap trick within the vectorized formalism, we can rewrite this expression making explicit the average over the noise and the unitaries:
\begin{equation}
\label{eq:fid_stat_mech}
    F=\llangle s|\prod_{i=1}^t \left\{ (\Noise_i\otimes \mathbbm{1})\Ex_{U_i}\left[(U_i\otimes U_i^*)^{\otimes 2} \right]\right\}\ket{\pmb{0},\pmb{0}}^{\otimes 2}\,,
\end{equation}
where $U_i$ denotes collectively the unitaries of the layer $i$. As in the previous section, Eq.~\eqref{eq:fid_stat_mech} defines an Ising-like statistical-mechanical model. In the large-time limit, the two dominant contributions come from the configurations with $e$ and $s$ permutation states everywhere, namely
\begin{equation}   \label{eq:fidelity_dw}F=\includegraphics[width=0.3\linewidth,valign=c]{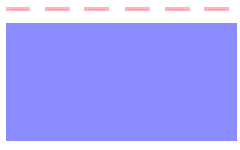}
    +\includegraphics[width=0.3\linewidth,valign=c]{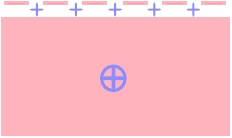}
\end{equation}
The left contribution ($e$ everywhere) evaluates to $d^{-N}$, as this is the cost of the top interface between the $e$ bulk and the $s$ boundary condition. The right contribution ($s$ everywhere) instead depends on the effect of noise in the circuit bulk, which we represent by a circled plus sign. We shall show below that, by defining $\Tilde{F}=F-d^{-N}$, the quantity 
\begin{equation}\label{eq:smallf}
    \fidel= - \frac{2}{N} \log_d \Tilde{F} 
\end{equation}
provides an effective measure of the number of errors per qudit in the circuit and generalizes the Hashing bound to this setting. A higher replica generalization of this quantity is discussed in End Matter. We expect a transition in the coherent information as a function of the parameter $\fidel$. The analog of the statistical-mechanical model in Eq.~\eqref{eq:statistical_model} for the two averaged purities appearing in Eq.~\eqref{eq:annealed_average} reads as
\begin{equation}
\begin{split}
     &\Ex_{U}\Tr(\rho_{B / RB}^\nreplicas)=\\ &=\llangle s|\prod_{i=1}^{t}\Noise_i^{\otimes 2}\Ex_{U_i}\left[(U_i\otimes U_i^*)^{\otimes 2}\right]|v_{B/RB}\rrangle\otimes |\pmb{0},\pmb{0}\rrangle^{\otimes 2}\,,
\end{split}
\end{equation}
where $\ket{\pmb{0}}=\ket{0}^{\otimes (N-k)}$ and $|v_{R/RB}\rrangle$ are defined before Eq.~\eqref{eq:statistical_model}. We now reexamine the purity of 
$B$ as this quantity is expected to capture the transition marking the end of the recovery phase. 
In the large-time limit, we have
\begin{equation}
    \Ex_{U} \Tr(\rho_B^2)=\includegraphics[width=0.3\linewidth,valign=c]{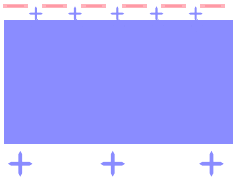}
    +\includegraphics[width=0.3\linewidth,valign=c]{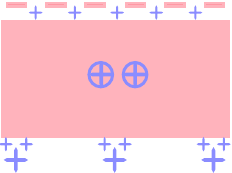}\,,
\label{eq:bulk_noise_purB}
\end{equation}
where the noise acting within the encoding circuit is represented by two circled plus signs, reflecting the fact that it appears in both replicas [in contrast to Eq.~\eqref{eq:fidelity_dw}, where it appears in only one]. Hence, the bulk contribution to the right term of Eq.~\eqref{eq:bulk_noise_purB} 
is the square of the right term in Eq.~\eqref{eq:fidelity_dw}, i.e.\ $\Tilde{F}^2$. We have however to incorporate the bottom and top boundary contributions, which are not included in the fidelity. These contributions are affected by noise, as depicted above by small plus signs. The bottom boundary terms are absent in the fidelity by construction. At the top boundary, the identity-to-swap contribution does not appear in the fidelity, since the latter includes noise in only a single replica. By contrast, the swap-to-swap contribution at the top boundary coincides with that entering the fidelity, at least to first order in the noise strength. Taken together, these contributions lead to
\begin{equation}
\label{eq:RM_bulk_purities}
    \begin{split}
        \Ex_{U}\Tr(\rho_B^\nreplicas)&=d^{-(g_{\swp,\idp}+1) N}+ d^{-[\fidel+r(1+g_{\swp,\idp})]N}\\
        \Ex_{U}\Tr(\rho_{RB}^\nreplicas)&=d^{-[g_{\swp,\idp}+1+r(g_{\idp,\swp}+1)]N}+d^{-(\fidel+g_{\swp,\swp} r) N}
        \,,%\\\Ex_{U}\Tr(\Lambda(\ket{\pmb{0}}\bra{\pmb{0}})^\nreplicas)&=d^{-(g_{\swp,\idp}+1) N}+\Tilde{F}^2d^{-g_{\swp,\swp}N}\,,
    \end{split}
\end{equation}
where $g_{\pi,\sigma}$, defined in Eq.~\eqref{eq:def_g_12}, represents the cost induced by the noise for the top and bottom boundary conditions in Eq.~\eqref{eq:bulk_noise_purB} and its analogous for the purity of $RB$. Equation~\eqref{eq:RM_bulk_purities} captures the error-correction transition at 
\begin{equation}
   [ \fidel]_{\rm critical}=(1-r)(1 + g_{\swp,\idp}).
\end{equation}
As errors accumulate throughout the circuit, $f_2$ grows linearly with depth $t$, whereas $g_{\swp,\idp}$ depends only on the per-gate noise rate $\gamma$. Maintaining a fixed $\fidel$ requires scaling $\gamma \sim 1/t$, which implies $g_{\swp,\idp} = O(1/t)$ (analogous scaling for $g_{\swp,\swp}$ and $g_{e,s}$). By contrast, if $f_2$ is allowed to grow with 
$t$, the critical noise rate at which the transition occurs decreases proportionally to 
$1/t$, ultimately vanishing at infinite depth. For weak single-qubit unital noise with rate $\gamma$, fidelity is given by $F \simeq e^{- \gamma N t}$, which implies $\gamma \leq  \log d (1-r)/(2 t)$, and $t=\omega(N)$ in order to achieve perfect recovery. For single-qudit depolarizing noise, $g_{\swp,\idp}=g_{\idp,\swp}=0$, we choose the noise strength to scale as $\gamma=\eta/t$, with $\eta$ independent of $t$, such that $f_2\approx 2\eta/\log d$ remains constant in time. This follows from the relation $F\approx e^{-\eta N}$. This implies that
$g_{\swp,\swp}=(d^2-1)f_2/(d^2 t)$
in the long-time limit. Substituting this expression into Eq.~\eqref{eq:RM_bulk_purities}, we obtain the leading correction in the recovery phase,
\begin{equation}
\label{eq:correction_setup_II_depo}
\frac{|\Ic-\Ic^\mathrm{RM}|}{rN}
=
\frac{d^2-1}{d^2}\frac{f_2}{t}.
\end{equation}

In Fig.~\ref{fig:info_depolarizing}(c), we show the coherent information for a brick-wall encoding circuit with depolarizing channels as a function of the global (log-of) fidelity $\fidel$. We observe a perfect recovery phase, followed by a transition away from it in the infinite-time limit. For a fixed fidelity, panel (d) shows the approach to the RM result, with $O(1/t)$ corrections.\\

\begin{table}[t]
\centering
\begin{tabular}{lcc}
\hline
 & \textbf{Setup I} & \textbf{Setup II} \\
\hline
\makecell{Noise Location} & After encoding & Within encoding \\ 
\makecell{Transition Location \\ at large depth} & $H_2=1-r$ & $f_2=1-r$ \\
\makecell{Perfect recovery \\ at large depth} & Yes & Only if $F=\mathrm{cst.}$ \\
\makecell{Finite depth correction \\ (recovery)} & $\propto N/e^{-t/2\tau}$ & $\propto N/t$ \\
\makecell{Finite depth correction \\ (transition)} & $\propto N/e^{-t/\tau}$ & $\propto N/t$\\
\bottomrule
\end{tabular}
\caption{Comparison between the two setups.}
\label{tab:comparison}

\end{table}

\section{Conclusion} We have studied error-correction transitions in one-dimensional noisy random quantum circuits. We focused on two different protocols: noise acting after the encoder (setup I) and noise acting within the encoder (setup II). Using a statistical-mechanical mapping, we have shown how in these two cases the coherent information approaches the same universal random matrix prediction at large circuit depth. Specifically, we find that a perfect error-protected phase is approached in parametrically different ways. While in setup I a circuit of logarithmic depth is enough, in the case of noisy encoder (setup II) the circuit depth should grow faster than linearly in system size $N$. For clarity, the main similarities and differences between setups I and II are summarized in Table~\ref{tab:comparison}, highlighting the key results of our work.
Overall, while short-depth encoding circuits are known to be achievable, at least for specific encoded states \cite{tan2025singleshotuniversalityquantumldpc, Hillmann_2025, nguyen2025quantumfaulttoleranceconstantspace}, our findings suggest that this does not extend to typical (average-case) scenarios, i.e.\ for random codes.
Looking ahead, a key direction is to extend the protocol beyond the case of incoherent noise considered here, in particular by incorporating measurements \cite{li2023statisticalmechanicsmonitoreddissipativerandomcircuits}
and feedback mechanisms, as well as coherent noise. Extensions to higher-dimensional encoders and to noisy decoding schemes are also left for future work.

Interestingly, we find that the same phenomenology and universality are displayed by the \textit{Holevo information} (see End Matter), recently studied in Ref.~\cite{Zhuang2023_DynamicalPhaseTransitions_InformationFlow_RandomCircuits}. This fact suggests the use of classical coherent information as a proxy for detecting information-protected phases of matter.\\

\begin{acknowledgments}
\section{Acknowledgments} 
We thank Yunos El Kaderi, Michael Gullans, Max McGinley, Michael Vasmer and Robert Koenig
for discussions. J.D.N., A.S., and G.L. are funded by the ERC Starting Grant Mo. 101042293 (HEPIQ) and the ANR-22-CPJ1-0021-01. 
P.R. acknowledges support by FCT through Grant No. UID/04540/2025 (DOI: 10.54499/UID/04540/2025), the I\&D unit Centro de Física e Engenharia de Materiais Avançados (CeFEMA), and Project SCALE-QLT (DOI: 10.54499/2024.16192.PEX).
This work was granted access to the HPC resources of IDRIS under the Allocations No. AD010613967R2 and No. AD010613967R1. This research was supported in part by Grant No. NSF PHY-2309135 to the Kavli Institute for Theoretical Physics (KITP).\\
\end{acknowledgments}

\section{Data availability}
The data that support the findings of this article are openly available \cite{dataavail}.

\bibliography{biblio}
\clearpage
\appendix 

\twocolumngrid
\begin{center}
    \textbf{\large End Matter}
\end{center}

\begin{figure}[ht!]
    \centering
    \includegraphics[width=1.0\linewidth]{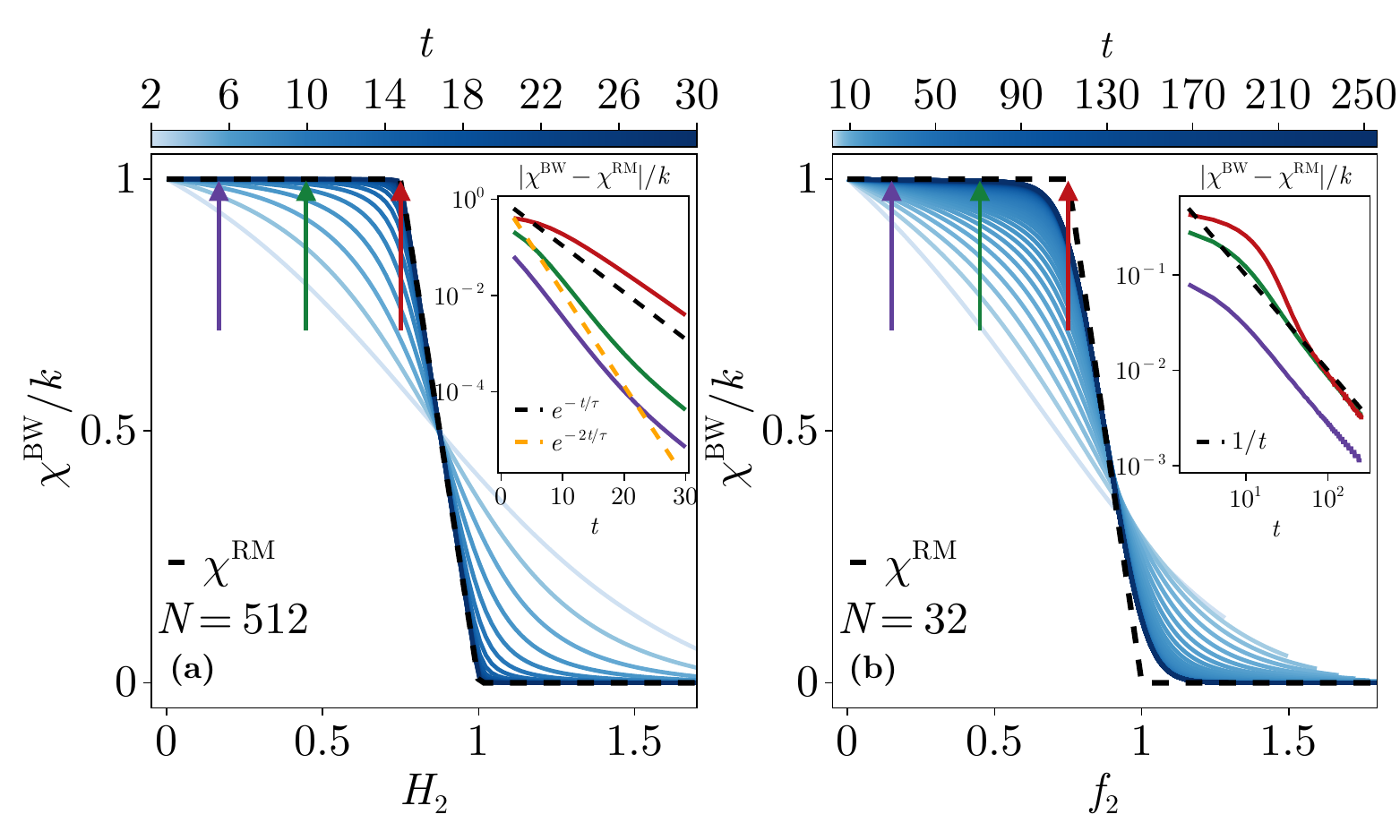}
    \caption{Setups I (a) and II (b) with depolarizing noise and encoding rate $r=1/4$. We show the Holevo information at different depths and noise strength and show the finite-time corrections to the RM result for three distinct levels of noise per qubits, two within the recovery regime and one at the transition point. While the general shape of the Holevo information is different than the coherent information, the critical noise strength at which the transition away from the recovery phase occurs and the scaling of the finite-time corrections are identical.}
    \label{fig:classical}
\end{figure}

\section{Classical information} While the coherent information quantifies the amount of quantum logical information that can be recovered, one may likewise characterize the amount of classical information accessible after the action of the noise. To this end, we disregard the reference system and focus solely on noise channel $\Lambda$. Let $\{\ket{\psi_i}\}_{i=1}^{d^N}$ denote the computational basis of the Hilbert space, and consider an ensemble in which these basis states are prepared with uniform probability. The corresponding accessible classical information is then quantified by the Holevo information \cite{Holevo1998,SchumacherWestmoreland1997}:
\begin{equation}
\label{eq:Holevo info}
    \Hol(\Lambda)=S\left(\sum_i p_i \Lambda(\ket{\psi_i}\bra{\psi_i})\right)-\sum_i p_i S\left[\Lambda(\ket{\psi_i}\bra{\psi_i})\right]\;,
\end{equation}
where $p_i=d^{-N}$. If we further assume that the channel is a Haar-random noisy circuit and restrict attention to annealed averages of second Rényi entropies, the averaged Holevo information reduces to
\begin{equation}
        \Hol=-\log_d\Ex_{U}\Tr\left[\Lambda_{U}(\mathbb{I}/d^N)^\nreplicas\right]+\log_d\Ex_{U}\Tr\left[\Lambda_{U}(\ket{\pmb{0}}\bra{\pmb{0}})^\nreplicas\right] \; ,
\end{equation}
where the first term reduces, upon application of the swap trick, to the purity of subsystem 
$B$ in the encoding picture: $\Tr\left[\Lambda_{U}(\mathbb{I}/d^N)^\nreplicas\right]=\Tr(\rho_B^2)$.\\

This observation implies that the analysis presented in the main text for the coherent information remains applicable to the Holevo information, up to a modification of the boundary conditions for one of the two terms in the information. In setup I, the large-$N$ limit of the classical information takes the form
\begin{equation}
    \Hol/N= \min( g_{\swp,\idp}+1,g_{\swp,\swp}+r)-\min(g_{\swp,\idp}+1,g_{\swp,\swp})\, .
\end{equation}
This expression is analogous to Eq.~\eqref{eq:RM info} and observes the same Hashing bound as the coherent information. Similarly, in setup II, there is also a transition as a function of $f_2=-2/N \log\Tilde{F}$.

In Fig.~\ref{fig:classical}, we display the infinite-time limit of the Holevo information together with its finite-time corrections in both setups and observe the same scaling as for the coherent information. \\

\begin{figure}[t!]
    \centering
    \includegraphics[width=1.0\linewidth]{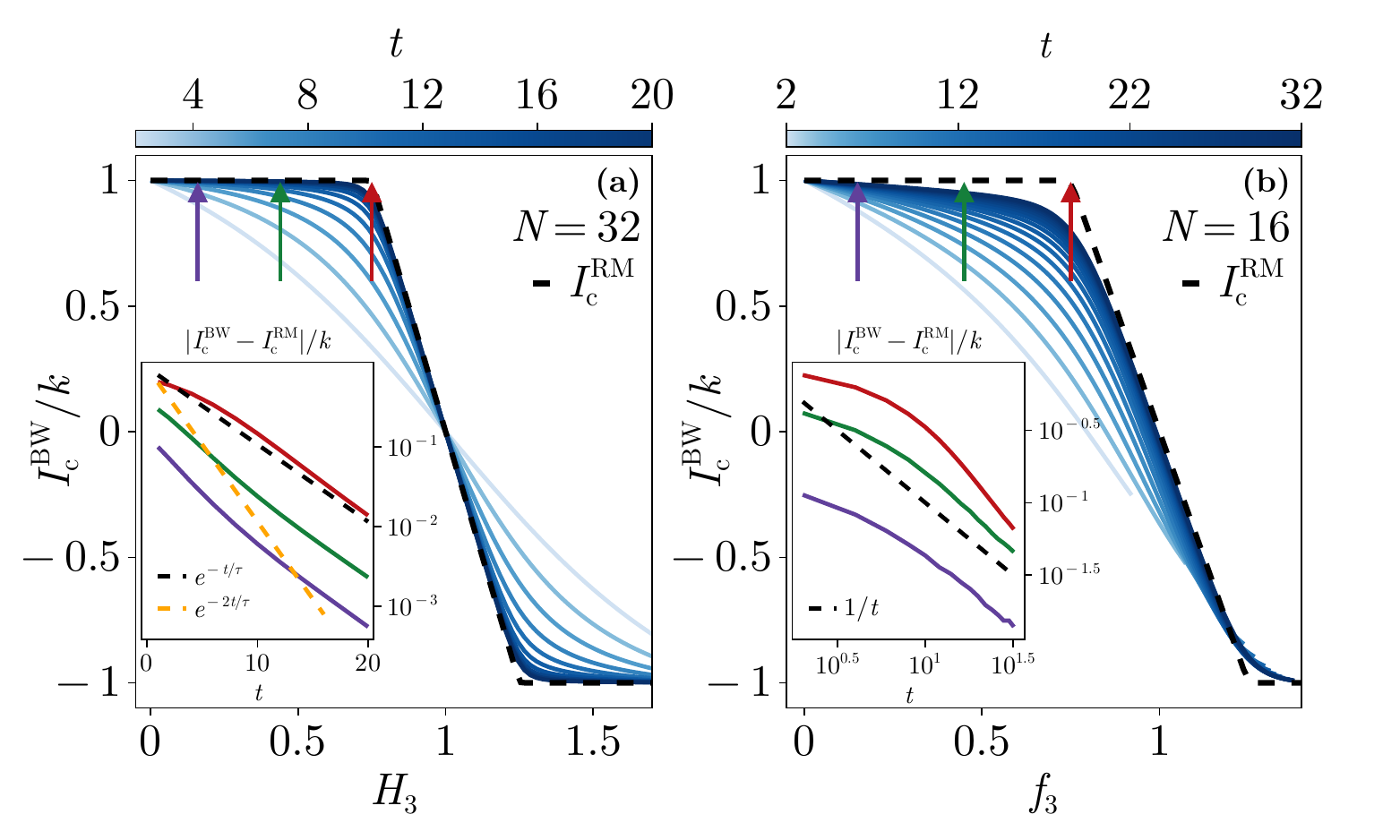}
    \caption{ Setups I (a) and II (b) with an encoding rate of $r=1/4$ and depolarizing noise where the coherent information now involves $3$-Rényi entropies. In both cases, we show that the corrections to the RM result are scaling in the exact same way as for $2$-Rényi entropies. The Hashing bound $H_3$ is defined in Eq.~\eqref{eq:depo_hashing_bound} while $f_3$ is defined in Eq.~\eqref{eq:f2higherreplica}.}
    \label{fig:info_3_replicas}
\end{figure}

\section{Higher number of replicas}
For setup I, computing the $\alpha$-Rényi entropies of $B$ and $RB$ requires evaluating overlaps of the form $\llangle \swp|\Noise^{\otimes \alpha}|\sigma\rrangle$, where $\swp$ now denotes the cyclic permutation $(2\; 3\;\dots\; \alpha \;1)$ and $\alpha>1$ is an integer. The behavior of these overlaps depends sensitively on the specific structure of the noise channel. In the infinite-time limit, it remains challenging to determine which permutation provides the dominant contribution at large system size $N$ when evaluating the coherent information. In the case of depolarizing noise, however, it becomes evident that the dominant permutations in the large-$N$ limit are the identity $\idp$ and the cycle $\swp$. This observation makes it possible to derive the Hashing bound for an arbitrary number of replicas \cite{Sommers2025}
\begin{equation}
    H_\alpha(\gamma)=\frac{1}{1-\alpha}\log_d\left[\left(1-\frac{d^2-1}{d^2}\gamma\right)^\alpha+(d^2-1)\left(\frac{\gamma}{d^2}\right)^\alpha\right].
    \label{eq:depo_hashing_bound}
\end{equation}
This observation further implies that the finite-time corrections to the RM result decay in the same manner for any 
$\alpha$, as illustrated in Fig.~\ref{fig:info_3_replicas}(a).\\

For setup II, we consider the weak-noise limit in which the local noise strength within the circuit scales as 
$\gamma=\eta/t$, with 
$\eta$ held fixed as a function of time and system size. In this regime, the effective noise contribution within the circuit can be expressed explicitly for arbitrary integer 
$\alpha>1$ (see \cite{Saulierenoise2025}). The two 
$\alpha$-Rényi moments entering the coherent information can then be written in closed form, temporarily neglecting noise contributions at the boundaries,
\begin{equation}
\label{eq:alpha coherent}
    \begin{split}
        \Ex_{U}\Tr(\rho_B^\alpha)&=d^{-\alpha(1+r)N}\sum_{\sigma\in S_\alpha}\left[\llangle s|\sigma\rrangle_d \Tilde{F}^{ \left[\alpha -n_\mathrm{F}(\sigma)\right]/N}\llangle \sigma|e\rrangle_d^{r}\right]^N\\
        \Ex_{U}\Tr(\rho_{RB}^\alpha)&=d^{-\alpha(1+r)N}\sum_{\sigma\in S_\alpha}\left[\llangle s|\sigma\rrangle_d^{1+r} \Tilde{F}^{ \left[\alpha -n_\mathrm{F}(\sigma)\right]/N}\right]^N
        \,,
    \end{split}
\end{equation}
where 
$n_\mathrm{F}(\sigma)$ denotes the number of fixed points of the permutation 
$\sigma$. By evaluating these expressions numerically, we find that it is sufficient to retain only the contributions from 
$e$ and $s$ in order to obtain an accurate approximation of the coherent information. Under this approximation, the coherent information exhibits a transition at
\begin{equation}\label{eq:f2higherreplica}
    f_\alpha:=-\frac{\alpha}{(\alpha-1)N}\log_d\Tilde{F}=1-r
\end{equation}
In Fig.~\ref{fig:info_3_replicas}(b), we display the transition in the three-replica coherent information as a function of 
$f_3$ and show that the finite-time corrections exhibit the same scaling as in the two-replica case. \\

\begin{figure}[t!]
    \centering
    \includegraphics[width=1.0\linewidth]{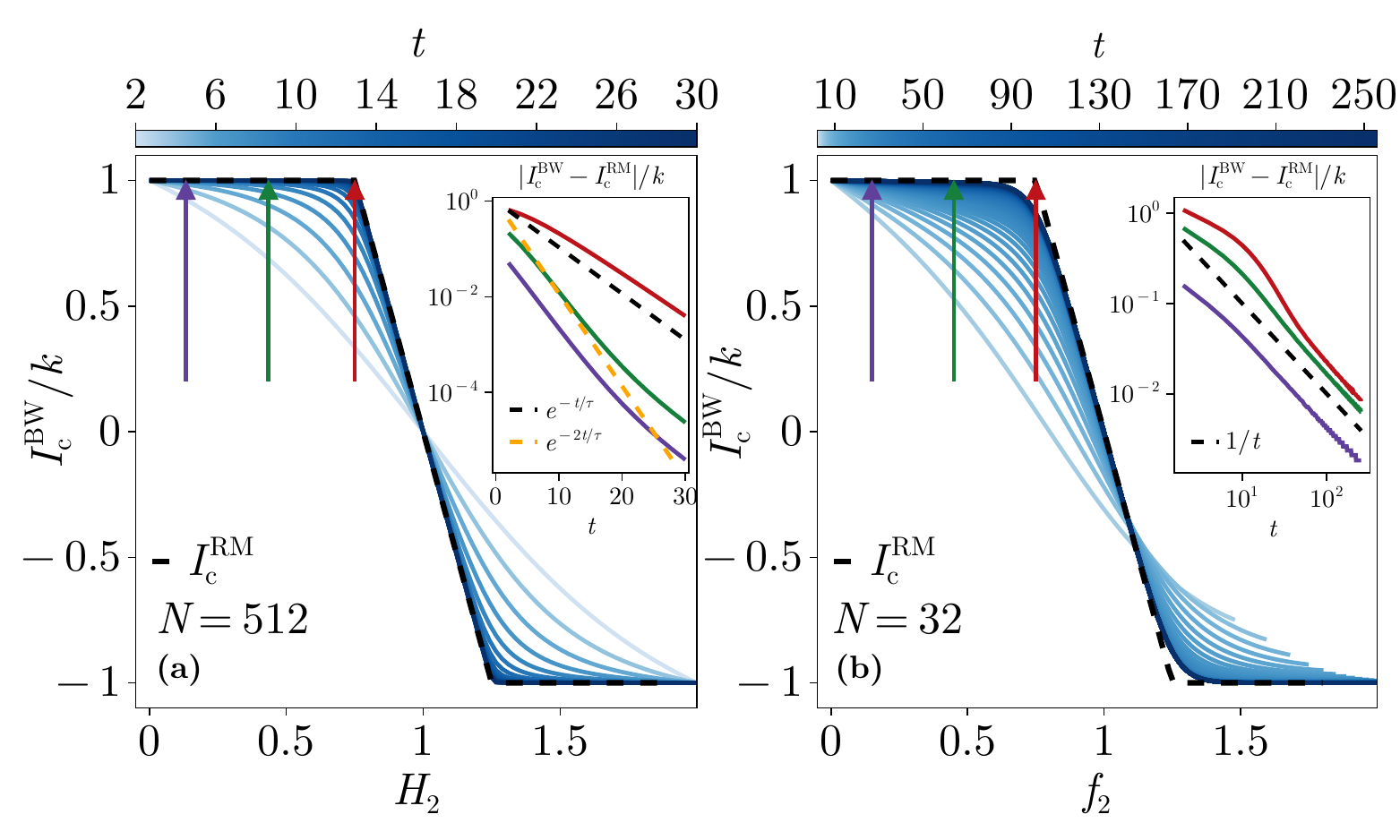}
    \caption{ Setups I (a) and II (b) with an encoding rate of $r=1/4$ and amplitude damping noise. In both cases, we show that the corrections to the RM result are scaling in the exact same way as the depolarizing noise (see Fig.~\ref{fig:info_depolarizing}). }
    \label{fig:info_ampdamp}
\end{figure}
\section{Amplitude damping}
In Fig.~\ref{fig:info_ampdamp}, we show the coherent information at finite depth for both setups, but now with amplitude damping noise $\Noise(\rho)=K_0\rho K_0^\dag + K_1\rho K_1^\dag$, where

\begin{equation}
    K_0=\begin{pmatrix}
        1 & 0 \\ 0 & \sqrt{1-\gamma}
    \end{pmatrix},\, K_1=\begin{pmatrix}
        0 & \sqrt{\gamma} \\ 0 & 0
    \end{pmatrix}
\end{equation}
are the Kraus operators of the channel. We find that the finite-time corrections are the same as for depolarizing noise.\\

\section{State design from Frame potential}

\begin{figure}[t!]
    \centering
    \includegraphics[width=0.7\linewidth]{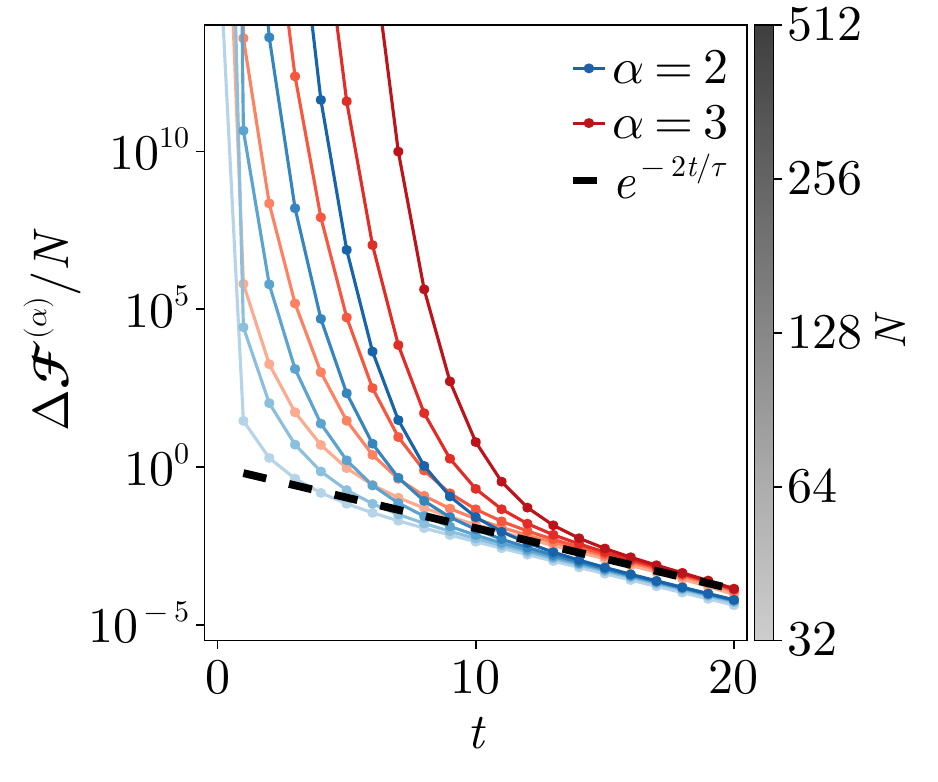}
    \caption{Plot of the distance of the average frame potential from the Haar value for a noiseless random brick-wall  circuit for different system sizes and depths, for $\alpha=2,\,3$, showing the approach to Haar with the slope $2/\tau$. }
    \label{fig:frame_potential}
\end{figure}
For an ensemble of states $\mathcal{E}$, we define the average frame potential,
\begin{equation}
\mathcal{F}^{(\alpha)}_\mathcal{E}=\Ex_{\psi,\psi'\in\; \mathcal{E}}\left[|\braket{\psi'|\psi}|^{2\alpha}\right]\;.
\end{equation}

An ensemble 
$\mathcal{E}$ is said to form a design if, for a given 
$\alpha$, its frame potential coincides with that of the Haar ensemble, namely $\mathcal{F}^{(\alpha)}_\mathrm{Haar}=D^{-2\alpha}\alpha!$, where $D$ denotes the Hilbert space dimension. It is known that Haar-random brick-wall circuits converge to designs in the limit of infinite circuit depth. A natural question is therefore how rapidly brick-wall circuits approach this limit as a function of time (or circuit depth). To quantify the deviation from the Haar value, we introduce
\begin{equation}
    \Delta\mathcal{F}^{(\alpha)}_\mathcal{E}=\frac{\mathcal{F}^{(\alpha)}_\mathcal{E}}{\mathcal{F}^{(\alpha)}_\mathrm{Haar}}-1\;.
\end{equation}
As shown in Fig.~\ref{fig:frame_potential}, for finite system size and finite-depth circuits the correction to the Haar frame potential scales as 
$N \, e^{-2t/\tau}$. Notably, this exhibits the same scaling behavior as the corrections observed in setup I of our analysis.\\

\section{Hashing bound} 
We want to express explicitly the quantity $H_2=g_{\swp,\swp}-g_{\swp,\idp}$ in the case of single-qubit Pauli channels $\Noise(\rho) = p_0 \Id + p_1 \PauliX \rho \PauliX + p_2 \PauliY \rho \PauliY + p_3 \PauliZ \rho \PauliZ$. Since this family of noise is unital, we have immediately that $g_{\swp,\idp}=0$. Using the definition of Eq.~\eqref{eq:def_g_12} and expanding the swap operator, we have 
\begin{equation}
    H_2=2-\log_2 \sum_{i,j}\Tr\left[\Noise(\ket{i}\bra{j})\Noise(\ket{j}\bra{i})\right]\,.
\end{equation}

Using $\Tr\left[\Noise(\ket{i}\bra{i})\Noise(\ket{i}\bra{i})\right]=(p_0+p_3)^2+(p_1-p_2)^2$ and $\Tr\left[\Noise(\ket{0}\bra{1})\Noise(\ket{1}\bra{0})\right]=(p_0-p_3)^2+(p_1+p_2)^2$, we find
\begin{equation}
    H_2=-\log_2(p_0^2+p_1^2+p_2^2+p_3^2)\,,
\end{equation}
which is exactly the $2$-Rényi entropy of the probability vector $(p_0, p_1, p_2, p_3)$, also known as the Hashing bound.
 
Finally, we remark, that for depolarizing noise, we have $H_2(\gamma)=2-\log_d\left[1+(d^2-1)(1-\gamma)^2\right]$ and, for amplitude damping noise, the Hashing bound is $H_2(\gamma)=-\log_d\left[1-(2-\gamma)\gamma/2\right]+\log_d\left[1+\gamma^2\right]$.
\end{document}